\documentclass[10pt,twocolumn,letterpaper]{article}

\usepackage[T1]{fontenc}
\usepackage{times}
\usepackage[letterpaper,textwidth=6.875in,textheight=8.875in,
            top=1in,columnsep=0.3125in,centering]{geometry}
\usepackage{amsmath,amssymb}
\usepackage{graphicx}
\usepackage{booktabs}
\usepackage{multirow}
\usepackage[font=small,labelfont=bf,labelsep=period,skip=4pt]{caption}
\usepackage{titlesec}
\usepackage[numbers,sort&compress]{natbib}
\usepackage[pagebackref,breaklinks,hidelinks]{hyperref}
\usepackage{url}

\titleformat{\section}{\fontsize{12}{14}\selectfont\bfseries}{\thesection.}{0.5em}{}
\titleformat{\subsection}{\fontsize{11}{13}\selectfont\bfseries}{\thesubsection.}{0.5em}{}
\titlespacing*{\section}{0pt}{10pt plus 2pt minus 2pt}{6pt plus 1pt}
\titlespacing*{\subsection}{0pt}{8pt plus 2pt minus 2pt}{4pt plus 1pt}
\titleformat{\paragraph}[runin]{\normalsize\bfseries}{}{0pt}{}
\titlespacing*{\paragraph}{0pt}{6pt plus 2pt minus 1pt}{0.5em}

\renewcommand*{\backref}[1]{}

\hypersetup{
  pdftitle={EWMA Principal Eigenportfolios of Equity Returns: Market Representation and Beta-Hedging Performance},
  pdfauthor={Andrew Kangmin Kim; Aurora Neyan Zhou}
}

\begin{document}

\twocolumn[{%
\begin{center}
{\Large\bfseries EWMA Principal Eigenportfolios of Equity Returns:\\[2pt]
Market Representation and Beta-Hedging Performance\par}
\vspace{16pt}
{\large Andrew Kangmin Kim$^{1}$ \qquad Aurora Neyan Zhou$^{2}$\\[2pt]
University of Oxford$^{1}$ \qquad University of Warwick$^{2}$\\[2pt]
{\normalsize\ttfamily kangmin.kim@hertford.ox.ac.uk \qquad AURORAZHOU12138@outlook.com}\par}
\vspace{20pt}
\end{center}
}]

\begin{center}
{\large\bfseries Abstract}
\end{center}
\vspace{-4pt}
{\itshape
This study examines whether an adaptively estimated principal eigenportfolio improves equity beta hedging relative to capitalisation-weighted and equal-weighted market benchmarks. Using stock returns from a historically changing S\&P 500 universe, it constructs a portfolio from the leading principal component of an exponentially weighted second-moment matrix of volatility-standardised returns. The resulting portfolio, EWMA-PC1, is compared with SPY and RSP under a common beta-hedging procedure across randomly generated portfolios of varying sizes and weighting schemes. Evaluation considers variance reduction and average pairwise residual correlation, with paired block-bootstrap inference. EWMA-PC1 exhibits broad market participation and achieves greater mean variance reduction and lower average residual correlation than SPY across all tested configurations. Its incremental advantage over RSP is smaller and mixed. For 50-stock Dirichlet portfolios, mean variance reduction is 92.17\% under PC1, compared with 79.03\% under SPY and 91.93\% under RSP; RSP instead achieves greater mean variance reduction for multi-stock liquidity-tilted portfolios. Statistical evidence of improvement is more consistent against SPY than against RSP. These findings distinguish the extraction of a market-like component from incremental hedge effectiveness and highlight the importance of benchmark weighting and target-portfolio composition. The results concern gross statistical risk reduction and exclude additional trading, financing, and borrowing costs.\par}

\vspace{6pt}

%% ======================================================================
\section{Introduction}\label{sec:intro}

Stock returns reflect both company-specific developments and influences shared across the market. This shared variation, or market comovement, means that holding more stocks can reduce idiosyncratic risk while leaving substantial exposure to broad market fluctuations. Investors seeking to retain selected stock positions while reducing this exposure therefore need an instrument whose returns capture the movement they wish to hedge. Broad equity index funds provide natural candidates, but their weighting rules determine which exposures they represent. Capitalisation weighting gives larger companies greater influence, whereas equal weighting distributes allocations evenly across constituents. Neither rule explicitly selects weights from the dependence observed among stock returns. This motivates a data-driven alternative: estimate a common return direction from those relationships, construct a corresponding portfolio, and test whether it provides a more effective market hedge.

Principal component analysis (PCA) offers an established foundation for this approach. Research on equity correlation matrices associates the leading component with broad market movement~\citep{plerou2002}, while eigenportfolio research translates statistical components into portfolio allocations and examines their interpretation as market proxies~\citep{avellaneda2022}. PCA-derived factors have also been compared with ETF-based factors in statistical-arbitrage strategies~\citep{avellanedalee2010}. Exponential weighting provides a way to update the estimated return relationships as new observations arrive~\citep{riskmetrics1996}. These foundations motivate an adaptive principal eigenportfolio, but its market resemblance does not by itself establish its usefulness as a hedge. Nor does the performance of a residual-based trading strategy directly establish how much risk a single market instrument removes. The question addressed here is whether an adaptively estimated principal eigenportfolio offers incremental beta-hedging effectiveness over both capitalisation-weighted and equal-weighted benchmarks, and how that comparison varies with the portfolio being hedged.

This study constructs an exponentially weighted moving average principal-component portfolio, denoted EWMA-PC1, from volatility-standardised stock returns in a historically changing S\&P 500 universe. The leading component is converted into stock weights with unit net investment, and its market characteristics are assessed through broad participation, explained variation, and association with benchmark returns. Its hedging effectiveness is then compared with SPY and RSP using the same beta-estimation procedure on randomly generated portfolios of different sizes under Dirichlet and liquidity-tilted weighting. Evaluation considers both individual portfolio variance reduction and average correlation among hedge residuals, with paired block-bootstrap comparisons assessing statistical uncertainty. PC1 exhibits broad market characteristics and achieves greater mean variance reduction and lower average residual correlation than SPY across the tested configurations. Its comparison with RSP is smaller and mixed: PC1 slightly improves mean variance reduction for Dirichlet portfolios, whereas RSP performs better for multi-stock liquidity-tilted portfolios. Statistical support is also more consistent against SPY than against RSP. The contribution is a comparative assessment of where statistical market extraction adds hedge value. The findings show why an equal-weight benchmark and variation in target-portfolio composition matter when interpreting an apparent advantage over a capitalisation-weighted index. They are consistent with weighting and exposure differences contributing to that advantage, without isolating their causal roles. The evaluation concerns gross statistical hedge effectiveness and excludes additional trading, financing, and borrowing costs.

The remainder of the paper is organised as follows. Section~\ref{sec:lit} reviews the relevant literature and develops the comparative research question. Section~\ref{sec:method} describes the data, principal-component portfolio construction, and hedging experiment. Section~\ref{sec:results} presents the component's empirical characteristics and the comparative hedging results. Section~\ref{sec:discussion} discusses their implications, limitations, and conclusions.

%% ======================================================================
\section{Literature review}\label{sec:lit}

\subsection{Market risk and the choice of market proxy}

Equity portfolios are exposed to common fluctuations that cannot be eliminated simply by increasing the number of holdings. The distinction between systematic and diversifiable risk provides the economic motivation for measuring market exposure and constructing a corresponding hedge. \citet{sharpe1964} places market risk at the centre of an equilibrium theory of expected returns. For hedge evaluation, however, the immediate question is how effectively an instrument offsets realised portfolio fluctuations. Evidence that a hedge reduces variance does not, by itself, test the expected-return restrictions of the capital asset pricing model.

The choice of an observable market proxy is therefore consequential. \citet{roll1977} distinguishes the theoretical market portfolio from the proxies used in empirical asset-pricing tests. Although his critique concerns the testability of asset-pricing theory, the distinction is also relevant when interpreting an equity-market hedge: an observable index represents a particular investment universe and weighting rule. A capitalisation-weighted index, an equal-weighted index and a statistically constructed portfolio can each reflect broad market movements without being interchangeable measures of exposure.

This distinction separates market representation from hedge effectiveness. A component may capture widespread co-movement and correlate strongly with established benchmarks, yet offer little incremental risk reduction for a particular portfolio. Conversely, differences in the composition of the target portfolio can alter the relative usefulness of alternative hedges. Evaluating a statistical market component therefore requires examining both its market characteristics and its ability to remove risk relative to observable alternatives.

\subsection{Principal components and the common equity-market mode}

Principal component analysis (PCA) provides a statistical approach to identifying common directions in a panel of equity returns. The leading component is the direction associated with the largest eigenvalue of the matrix being decomposed. Its interpretation depends on the inputs: applying PCA to raw-return covariance gives greater influence to differences in return scale, whereas standardising returns changes the objective towards common variation across comparably scaled series. \citet{avellaneda2020} describes this standardised-return framework and its connection to equity risk factors. A component extracted in this space need not have the same composition as a capitalisation-weighted market index.

Research on financial correlation matrices provides an empirical basis for interpreting the leading component as a market mode. \citet{laloux1999} show that much of the observed eigenvalue spectrum can resemble that generated by random matrices, highlighting the difficulty of separating information from estimation noise. \citet{plerou2002} associate the largest eigenvalue with an influence common to the stock universe, while other informative eigenvectors reveal groups corresponding to business sectors. Together, these findings support distinguishing a broad common component from both weaker structured variation and noise; they do not imply that a single component captures all economically relevant equity risk.

Broad participation, a substantial explained share and strong association with benchmark returns provide evidence for a market interpretation, rather than identification of a unique causal factor. The fraction of matrix variation associated with the leading eigenvalue describes the input representation, whereas the fraction of a target portfolio's variance removed by hedging depends on its relationship with the constructed hedge.

\subsection{From principal components to equity eigenportfolios}

Turning a principal component into a stock portfolio requires distinguishing coefficients on standardised returns from weights on ordinary returns. If a component combines each stock's return after division by its volatility, the corresponding raw-return coefficient is the eigenvector entry divided by that volatility. A further normalisation can impose an investment budget. This conversion links a statistical direction to a portfolio return that can be compared with observable market instruments.

\citet{avellanedalee2010} provide an important precedent for using PCA-derived factors alongside ETF-based factors in US equities. Their framework models residual movements and uses them to construct statistical-arbitrage strategies. It establishes a connection between statistical factor extraction, equity portfolios and ETF alternatives. Its trading objective differs from evaluating the risk removed by a single market hedge: residual-based strategy performance also depends on the modelling of residual behaviour and the trading signals built from it.

A particularly close predecessor is \citet{avellaneda2022}, who examine principal eigenportfolios as proxies for the equity market. Under a large-dimensional model, they relate the leading correlation eigenvector to market betas divided by stock return standard deviations. Their construction includes conversion by inverse volatility and normalisation to unit net investment, and they compare conventional and hierarchical eigenportfolios with a capitalisation-weighted market portfolio. Their analysis provides direct precedent for both the market interpretation of a principal eigenportfolio and its construction from standardised-return loadings.

Related work considers how additional structure affects the extracted factors. \citet{avellaneda2020} incorporates sector groupings through hierarchical PCA. More recently, \citet{mudalgikar2024} investigate a value-informed construction and report improvements in eigenportfolio stability. These approaches illustrate that the organisation of the input data can influence the resulting portfolio. Evidence about market correspondence or stability nevertheless leaves a further practical question: whether the resulting portfolio offers incremental hedge effectiveness relative to conventional market instruments.

\subsection{Time-varying volatility and covariance estimation}

Estimating a common return direction requires choosing how historical observations inform current risk. Exponential weighting provides a recursive alternative to an equally weighted rolling window: recent observations receive greater weight, while the contribution of older observations declines gradually. RiskMetrics~\citep{riskmetrics1996} is an established reference for this approach and for treating daily conditional mean returns as zero when estimating second moments.

Separating marginal volatility from dependence is also useful conceptually. \citet{engle2002} develops dynamic conditional correlation models that combine individual volatility models with a parsimonious representation of changing correlations. This distinction motivates treating return scales and co-movement as separate estimation problems, potentially with different degrees of persistence. It does not make every matrix constructed from volatility-standardised returns a dynamic conditional correlation model; the specification of the matrix and its normalisation remain consequential.

Adaptation does not remove estimation uncertainty. \citet{ledoitwolf2004} demonstrate the relevance of covariance estimation error for portfolio applications and develop shrinkage as a response. Shrinkage and exponential weighting address different aspects of the estimation problem: the former regularises the matrix estimate, while the latter changes the relative influence of observations through time. Exponential weighting therefore provides a practical way to accommodate changing relationships, rather than a guarantee of noise-free factor estimates or optimal memory parameters.

\subsection{Portfolio weighting and the choice of hedge benchmark}

Weighting rules alter the exposures represented by an equity portfolio. In their working paper, \citet{plyakha2012} compare equal-, value- and price-weighted portfolios and relate performance differences to systematic exposures and rebalancing. Their analysis concerns investment performance rather than hedge effectiveness, but it demonstrates why portfolios drawn from a common stock universe need not represent the same risks. A comparison between a broad statistical component and a capitalisation-weighted index may therefore reflect allocation differences as well as the method used to estimate the component.

Simple allocation rules also provide demanding empirical benchmarks. \citet{demiguel2009} find that the estimation-based portfolio strategies they evaluate do not consistently outperform a naive $1/N$ allocation out of sample across their performance criteria and datasets. Their results do not establish how an equal-weighted ETF will perform as a hedge. They do support evaluating an estimated portfolio against a simple alternative before attributing practical value to its additional complexity.

Including RSP alongside SPY follows this benchmarking logic. SPY represents capitalisation-weighted exposure, while RSP provides an equal-weighted comparison. An advantage over SPY that does not extend to RSP would support a narrower conclusion than a general advantage of statistical extraction over conventional proxies. At the same time, comparing PC1 with RSP does not isolate weighting as a causal mechanism: the instruments differ in their precise allocations, updating and implementation.

The allocation of the target portfolio is equally relevant. A hedge that closely matches one set of stock exposures may leave more residual risk for another. Evaluating different portfolio sizes and allocation patterns therefore helps determine whether a hedge's usefulness extends beyond a particular target portfolio. A common set of targets also makes comparisons between hedge instruments more informative than evaluations conducted on different underlying holdings.

\subsection{Evaluating hedge effectiveness and remaining common risk}

The minimum-variance approach evaluates a hedge by the risk remaining after its return is subtracted in an appropriate proportion from the target return. \citet{ederington1979} is a classical reference for assessing hedge performance through variance reduction. For a single hedge instrument with positive variance, minimising the variance of the hedged return gives a coefficient equal to target-hedge covariance divided by hedge variance. Its realised effectiveness depends on how the required moments are estimated and how exposures evolve.

Individual variance reduction does not fully describe the common risk remaining across portfolios. \citet{famafrench1993} identify shared equity-return variation associated with factors beyond the market, providing a reason why a single market hedge need not remove all co-movement. More directly, \citet{avellaneda2020} notes that residuals from a one-factor representation can remain correlated with one another even when they are uncorrelated with the fitted factor. Shared holdings can also contribute to dependence among portfolios. These considerations motivate examining remaining co-movement as well as the risk removed from each portfolio.

Statistical comparisons must also account for dependence in the observations. \citet{politisromano1994} introduce the stationary bootstrap for dependent data, providing a basis for inference that retains local temporal relationships.

\subsection{Research gap and positioning of the present study}

The literature establishes market-like principal components, provides direct precedents for constructing eigenportfolios, and demonstrates their use in residual-based trading. It also shows why weighting and estimation choices can affect portfolio behaviour. These findings motivate a distinct comparative question: does an adaptively estimated principal eigenportfolio offer incremental beta-hedging effectiveness over both capitalisation-weighted and equal-weighted market benchmarks? Market correspondence and trading-strategy performance alone do not answer that question, particularly when the target portfolio's allocation affects the exposure to be removed.

The present study addresses this question by first establishing the empirical market characteristics of an EWMA-PC1 portfolio and then comparing PC1, SPY and RSP hedges on the same randomly generated portfolios. Varying portfolio size and allocation, and assessing both individual variance reduction and residual co-movement, provides evidence on where the candidate hedge adds value and where an existing market proxy performs comparably or better.

%% ======================================================================
\section{Methodology}\label{sec:method}

The study asks whether the leading principal component of volatility-standardised equity returns captures broad market co-movement and whether a portfolio constructed from that component provides an effective beta hedge relative to SPY and RSP. The first question is assessed through the component's participation, explained share and benchmark association. The second is assessed by applying each hedge to the same randomly generated stock portfolios and measuring both individual portfolio variance and residual co-movement across portfolios.

\subsection{Data and sample construction}\label{sec:data}

Daily adjusted closing prices are obtained from Yahoo Finance for historical S\&P 500 constituents and the SPY and RSP exchange-traded funds. SPY provides a capitalisation-weighted benchmark and RSP an equal-weighted benchmark. Historical membership is reconstructed by reversing dated additions and removals from cached Wikipedia constituent records. Membership changes are applied on their recorded dates; events after the sample end are also reversed when reconstructing earlier membership.

The price panel runs from 2 January 2015 to 21 August 2026 and contains 2,926 dates. The first 756 dates are reserved for estimator warm-up. Subsequent dates with available returns for PC1, SPY and RSP form the analysis window: 3 January 2018 to 21 August 2026, comprising 2,170 observations. Cached price and membership snapshots fix the inputs used in the reported run.

Of 744 historical tickers requested, 134 have no downloaded price history. Four further tickers are excluded by a listing-discontinuity screen: CEG, DOW, Q and SNDK. The screen flags membership gaps spanning at least 60 in-sample panel dates when fewer than 20\% have observed prices. Eight retained histories do not overlap the ticker's relevant membership. The resulting panel contains 606 stock columns, of which 598 contribute during membership. Historical reconstruction therefore improves universe coverage but does not eliminate selection arising from unavailable histories.

Simple returns are calculated between consecutive panel dates,
\begin{equation}
  r_{i,t} = P_{i,t} / P_{i,t-1} - 1. \label{eq:ret}
\end{equation}
A return is valid when both prices are observed and positive and the calendar gap is at most four days. Missing prices are not forward-filled. Let $V_{i,t}$ indicate return validity and $M_{i,t}$ historical index membership.

The four-day rule permits ordinary weekend and holiday gaps while excluding returns spanning longer gaps in the price panel. Membership and return availability have distinct roles: membership determines which stocks can enter the daily PC1 construction, while validity determines whether a stock return can be used on a given date. The test portfolios introduced below retain their original holdings after formation, subject to return availability.

\subsection{EWMA principal-component estimation}\label{sec:ewma}

Scaling each stock's return by its lagged volatility places returns on comparable scales, so differences in individual volatility alone do not dominate the estimated common direction. Using the lagged scale avoids incorporating the return being standardised into its own volatility estimate. For each valid observation, the stock's second moment is updated as
\begin{equation}
  s_{i,t} = \lambda_\sigma s_{i,t-1} + (1-\lambda_\sigma)\, r_{i,t}^2, \qquad \lambda_\sigma = 2^{-1/60}. \label{eq:vol}
\end{equation}
Here $\lambda_\sigma \approx 0.9885$, so an observation's contribution halves after 60 valid updates ($\lambda_\sigma^{60} = 1/2$). Daily returns are treated as having zero conditional mean for second-moment estimation, a convention also used in RiskMetrics~\citep[Section~5.3.1.1]{riskmetrics1996}. The square root of the estimated second moment provides the volatility scale.

The state starts at the first valid squared return and freezes when a return is unavailable. Volatility is $\sigma_{i,t-1} = \max(\sqrt{s_{i,t-1}},\, 10^{-4})$ and becomes usable after 20 prior valid observations. The decay thus advances on valid observations. Available pre-membership returns also contribute to the volatility estimate.

The active set $A_t$ contains current members with valid returns and available lagged volatility. Standardised matrix inputs are $\tilde z_{i,t} = \operatorname{clip}(r_{i,t}/\sigma_{i,t-1}, -10, 10)$ for active stocks and zero otherwise. Volatility and matrix estimates use prior returns; stocks enter the daily portfolio when their current-day return is available.

The full second-moment matrix starts at the identity. On a stock's entry or re-entry into the index, its row and column are reset to zero off-diagonal and one on the diagonal. Writing $R_t$ for this reset operation, the daily recursion is
\begin{equation}
\begin{aligned}
  S_{t-} &= R_t(S_{t-1}), \\
  S_t &= \lambda_S S_{t-} + (1-\lambda_S)\,\tilde z_t \tilde z_t^{\top}, \qquad \lambda_S = 2^{-1/252}.
\end{aligned} \label{eq:matrix}
\end{equation}
Here $\lambda_S \approx 0.9973$, giving the matrix a half-life of 252 panel dates. Clipping limits the influence of extreme standardised observations on matrix estimation; it does not cap the realised portfolio return.

The shorter volatility half-life and longer matrix half-life give the two estimates different degrees of persistence: individual return scales use a shorter memory, while the common direction is estimated from a longer history of co-movement. The half-lives of 60 valid volatility updates and 252 panel dates specify this responsiveness-stability trade-off; they are fixed settings of the reported specification rather than estimated optimal horizons.

The matrix decays on every input-panel date, including when some inputs are inactive. The same zero-mean approximation is used for matrix estimation, so standardised returns enter without sample-mean subtraction. Accordingly, $S$ is an uncentred second-moment estimate used to characterise return co-movement, not an estimator of expected-return drift. Its diagonal is not constrained to one. Entrant resets give new members a common initial prior. The 756-date warm-up spans three matrix half-lives; later entrants retain their own more recent initialisation.

Before today's matrix update, the leading eigenvector $v_t$ is extracted from $S_{t-}[A_t, A_t]$, provided at least 30 stocks are active. Its sign is aligned with the previous available vector on overlapping names, with a positive loading sum used when no reference is available. Portfolio weights and the realised return are calculated next; only then does today's clipped input update $S$.

\subsection{PC1 portfolio construction and characterisation}\label{sec:pc1}

The eigenvector supplies coefficients for extracting a component score from standardised returns. To express that same score as a weighted sum of ordinary stock returns, let $z_{i,t} = r_{i,t}/\sigma_{i,t-1}$ denote the unclipped standardised return. Then
\begin{equation*}
  f_t = \sum_{i \in A_t} v_{i,t} z_{i,t} = \sum_{i \in A_t} (v_{i,t}/\sigma_{i,t-1})\, r_{i,t}.
\end{equation*}
The stock-return coefficients $v_{i,t}/\sigma_{i,t-1}$ are normalised by their sum to give unit net investment:
\begin{equation}
  a_{i,t} = v_{i,t} / \sigma_{i,t-1}, \qquad w_{i,t} = a_{i,t} \Big/ \sum_{j \in A_t} a_{j,t}. \label{eq:weights}
\end{equation}
The eigenvector's arbitrary scale and sign cancel in this normalisation. Negative weights are retained, so unit net exposure does not impose unit gross exposure. Construction is skipped when $|\sum_i a_{i,t}| \le 10^{-3} \sum_i |a_{i,t}|$, guarding against near-cancelling coefficients.

This normalisation imposes $\sum_i w_{i,t} = 1$ in place of the eigenvector's usual $\sum_i w_{i,t}^2 = 1$, which does not represent an investment budget; it does not make the individual stock weights equal.
\begin{equation}
  r_{\mathrm{PC1},t} = \sum_{i \in A_t} w_{i,t} r_{i,t}. \label{eq:pc1ret}
\end{equation}
Weights are recomputed daily. Realised stock returns are neither clipped nor standardised, and the portfolio is not subsequently volatility-matched to either ETF. No additional trading, financing or borrowing costs are deducted; ETF returns retain expenses and tracking effects embedded in their adjusted-price series.

Market characteristics are assessed through the fraction of positive weights, active-name count, sum of the ten largest signed weights, gross short exposure and Pearson correlations with SPY and RSP. The explained share is $\lambda_{1,t}/\operatorname{tr}(S_{t-}[A_t, A_t])$, using the same matrix as the eigendecomposition. These measures describe broad market participation and captured second-moment variation.

The explained share is thus the proportion of estimated standardised second-moment variation associated with PC1, not the percentage of raw stock-return variance removed by the hedge. A cumulative log-return plot, $\sum_{s \le t} \log(1 + r_{H,s})$, provides a common visual comparison of the three return histories.

\subsection{Test-portfolio design}\label{sec:design}

Test portfolios are formed once on 3 January 2018. Eligible stocks must be index members with valid returns on that date and a positive, finite liquidity proxy, producing a universe of 430 names. For each portfolio size $n \in \{1, 10, 20, 50\}$, 1,000 sets of stocks are sampled uniformly without replacement within each set. Different portfolios may overlap. Each selected set receives two alternative allocations: symmetric Dirichlet weights and liquidity-tilted weights, defined below. The same name selections are used for these two schemes, and each resulting portfolio is evaluated against all three hedges.

Under the first scheme, weights follow $\mathrm{Dirichlet}(1, \dots, 1)$, giving nonnegative weights that sum to one. Each selected stock has expected weight $1/n$, but a particular draw can be concentrated and is generally not equally weighted. This scheme varies portfolio exposures without systematically favouring a name because of its liquidity. Under the liquidity-tilted scheme,
\begin{equation}
  u_i = q_i g_i \Big/ \sum_j q_j g_j, \qquad g_i \sim \mathrm{Exp}(1). \label{eq:liq}
\end{equation}
Here $q_i$ is the median available adjusted-price-times-volume observation over the 252 panel dates preceding formation. No further minimum observation count is imposed, and the proxy remains fixed thereafter. It is a liquidity proxy rather than actual dollar turnover or market capitalisation. The tilt changes both allocation and concentration. All draws use reproducible random streams with base seed zero.

Multiplying the positive random draw $g_i$ by $q_i$ shifts allocation toward more liquid names while preserving variation across portfolios. If all selected names had the same proxy, normalised exponential draws would have the same Dirichlet distribution as the first scheme. The two designs therefore assess hedge effectiveness under different portfolio exposures; the liquidity tilt does not replicate the weighting of SPY or any other index fund. With one stock, both schemes assign it weight one and give identical results.

Original target weights remain fixed, with daily renormalisation over valid holdings:
\begin{equation}
  r_{P,t} = \frac{\sum_i u_i V_{i,t} r_{i,t}}{\sum_i u_i V_{i,t}}. \label{eq:portret}
\end{equation}
A return is recorded only when at least 50\% of original weight remains valid. Stocks remain eligible holdings after leaving the index while their returns are available, and temporarily unavailable holdings resume contributing when valid returns return. This constructs constant-target-weight portfolios on available holdings; it does not reconstruct buy-and-hold wealth or liquidation proceeds for missing securities.

\subsection{Hedge construction}\label{sec:hedge}

Each portfolio is hedged separately against PC1, SPY and RSP. Stock--hedge covariances and hedge variance are estimated from raw daily returns using centred sample moments in equally weighted rolling windows of 126 analysis dates, requiring at least 60 observations. Covariances use available paired stock--hedge returns; hedge variance uses available hedge returns. The coefficient is refreshed every 21 analysis dates using estimates through the preceding date and held fixed between refreshes.

The 126-date window covers approximately six trading months, and refreshing every 21 analysis dates gives an approximately monthly schedule. These window and refresh settings are applied uniformly across PC1, SPY and RSP; they are not asserted to be optimal for any instrument.

At refresh date $\tau$, let $I_{i,H,\tau}$ contain the dates with available paired stock and hedge returns within the preceding 126 analysis dates, ending at $\tau - 1$, and let $m_{i,H,\tau}$ be its size. The covariance estimate is
\begin{equation}
\begin{aligned}
  &\widehat{\mathrm{Cov}}_{\tau-1}(r_i, r_H) \\
  &\quad = \frac{1}{m_{i,H,\tau} - 1} \sum_{s \in I_{i,H,\tau}} (r_{i,s} - \bar r_{i,I})(r_{H,s} - \bar r_{H,I}).
\end{aligned} \label{eq:cov}
\end{equation}
Here $\bar r_{i,I}$ and $\bar r_{H,I}$ are the respective sample means over the paired dates in $I_{i,H,\tau}$; the estimate requires $m_{i,H,\tau} \ge 60$. Let $J_{H,\tau}$ contain available hedge-return dates within the same window, with size $k_{H,\tau}$. Hedge variance is estimated as
\begin{equation}
  \widehat{\mathrm{Var}}_{\tau-1}(r_H) = \frac{1}{k_{H,\tau} - 1} \sum_{s \in J_{H,\tau}} (r_{H,s} - \bar r_{H,J})^2. \label{eq:var}
\end{equation}
The mean $\bar r_{H,J}$ is calculated over $J_{H,\tau}$, and $k_{H,\tau} \ge 60$ is required. When a stock has missing returns, its paired covariance sample can be smaller than the sample used for hedge variance.

The EWMA matrix estimates the common return direction used to construct PC1, while the centred rolling moments estimate stock exposures to each already constructed hedge instrument. This separates instrument construction from exposure estimation and compares PC1, SPY and RSP under a common hedge-estimation procedure. For portfolio return $r_P$ and hedge return $r_H$, minimising $\mathrm{Var}(r_P - b\, r_H)$ over $b$ gives $b = \mathrm{Cov}(r_P, r_H)/\mathrm{Var}(r_H)$, provided hedge variance is positive. This motivates the covariance-to-variance exposure ratio below, which combines estimated stock--hedge covariances using the portfolio's effective weights:
\begin{equation}
  \hat\beta_{P,H,\tau} = \frac{\sum_i \tilde u_{i,\tau-1,H}\, \widehat{\mathrm{Cov}}_{\tau-1}(r_i, r_H)}{\widehat{\mathrm{Var}}_{\tau-1}(r_H)}. \label{eq:beta}
\end{equation}
The weights $\tilde u$ renormalise original target weights over holdings with valid returns and available covariance estimates at $\tau - 1$, again requiring at least 50\% of original weight. This constituent-level exposure estimator need not equal direct rolling OLS on the realised portfolio when observation coverage or effective weights change.

With fixed portfolio weights and a common observation set, the identity $\mathrm{Cov}(\sum_i u_i r_i, r_H) = \sum_i u_i \mathrm{Cov}(r_i, r_H)$ motivates this aggregation. Computing stock-level covariances once permits the same estimates to be used across all portfolio draws.

The first refresh follows 126 analysis observations, making the first hedge active on 5 July 2018. The nominal hedge window contains 2,044 dates through 21 August 2026. Hedged returns, per unit original portfolio notional, are
\begin{equation}
  \varepsilon_{P,H,t} = r_{P,t} - \hat\beta_{P,H,t}\, r_{H,t}. \label{eq:resid}
\end{equation}
The coefficient is not updated using the return it is applied to. These residuals include no intercept subtraction, risk-free adjustment or separate cash-financing return.

\subsection{Performance measures and statistical inference}\label{sec:measures}

Hedge performance is assessed along two complementary dimensions: the variance remaining in individual portfolios and average co-movement across hedged portfolios. For each portfolio and instrument, variance reduction is
\begin{equation}
  \mathrm{VR}_{P,H} = 1 - \mathrm{Var}(\varepsilon_{P,H}) / \mathrm{Var}(r_P). \label{eq:vr}
\end{equation}
Both variances use the same valid dates for that portfolio and hedge, with at least 50 observations required. Negative values are retained. The main results report the mean across portfolios; the 10th and 90th percentiles describe cross-portfolio dispersion rather than uncertainty about that mean. The mean gives each eligible portfolio equal weight after computing its own proportional reduction; it is not obtained by pooling all residual returns into a single variance.

\paragraph{Paired comparisons of squared residuals.} Statistical comparison uses the same portfolio on the same date under two alternative hedges. For instruments A and B, the daily paired difference is
\begin{equation}
  d_t = |\mathcal{P}_t|^{-1} \sum_{P \in \mathcal{P}_t} \bigl(\varepsilon_{P,A,t}^2 - \varepsilon_{P,B,t}^2\bigr), \label{eq:dt}
\end{equation}
where $\mathcal{P}_t$ contains portfolios observed under both instruments. Averaging $d_t$ over dates yields the mean squared-residual difference. This is an absolute measure, distinct from average percentage variance reduction; it also includes any contribution from squared residual means.

Pairing holds the underlying investment portfolio and market date fixed, so the difference reflects the alternative hedge applied to that portfolio. A negative PC1~$-$~SPY or PC1~$-$~RSP estimate indicates smaller squared residuals under PC1. Because this statistic is in squared-return units rather than divided by each portfolio's original variance, its ranking can differ from mean percentage variance reduction.

\paragraph{Residual co-movement.} Cross-sectional co-movement is measured by mean off-diagonal residual correlation,
\begin{equation}
  \bar\rho_H = \frac{2}{K(K-1)} \sum_{j<k} \mathrm{Corr}(\varepsilon_{j,H}, \varepsilon_{k,H}). \label{eq:rhobar}
\end{equation}
Each correlation is calculated over time for a pair of portfolio residual series, and the result is averaged across all distinct pairs among $K$ portfolios. This measures how much average co-movement remains after hedging. It complements individual variance reduction because a hedge can reduce fluctuations in each portfolio while leaving a common pattern across portfolios. Shared holdings can also contribute to residual correlation, so zero is not assumed to be the attainable benchmark.

For each descriptive panel, all-missing dates are removed, portfolios with less than 90\% coverage are excluded, and up to 250 surviving portfolios are selected using seed zero. Only dates complete across the selected portfolios are retained. Unhedged correlations are calculated after restricting the panel to dates on which PC1 hedge residuals exist. These descriptive filters are applied separately by instrument.

Paired correlation comparisons instead use a common grid across all three hedges: portfolios must meet the coverage rule under every instrument, the same subset is selected, and retained dates must be complete for all panels. The resulting point differences and intervals are therefore paired, although they need not equal differences between separately filtered descriptive entries. Lower $\bar\rho$ denotes lower average signed correlation, not necessarily lower dependence by every measure.

\paragraph{Bootstrap confidence intervals and statistical significance.} Uncertainty is assessed using a stationary block bootstrap with geometrically distributed block lengths averaging 21 retained observations, circular continuation at the sequence boundary, 500 replications and 95\% percentile intervals. For squared residuals, the bootstrap resamples the daily $d_t$ series after removing missing dates. For correlation, identical resampled date indices are applied to both aligned panels and correlations are recomputed. A difference is called statistically significant when its interval excludes zero; an interval spanning zero does not establish equivalence.

Blocks preserve local temporal dependence instead of treating individual dates as independent draws. Resampling dates also keeps the contemporaneous cross-portfolio relationships intact: the 1,000 portfolios share the same market history and are not treated as 1,000 independent histories.

The bootstrap operates on realised performance histories: portfolios are not redrawn and PCA and hedge ratios are not re-estimated. The intervals describe temporal sampling uncertainty conditional on those histories and are unadjusted for multiple comparisons.

%% ======================================================================
\section{Results}\label{sec:results}

\subsection{Empirical characteristics of the PC1 portfolio}\label{sec:characteristics}

The PC1 construction produces a broad, almost entirely positive portfolio throughout the analysis window (Table~\ref{tab:characteristics}). The median fraction of positive weights is 100\%, and its minimum is 98.80\%. Median top-ten weight is only 4.06\%, while maximum gross short exposure is 0.0273\%. With unit net exposure, this implies maximum gross exposure of approximately 100.055\%; the observed portfolio is therefore close to fully long-invested rather than materially leveraged.

\begin{table}[t]
\centering
\caption{Characteristics of the PC1 portfolio. Analysis window: 3 January 2018--21 August 2026. Explained share refers to the active uncentred second-moment matrix; top-ten weight sums the ten largest signed weights. Correlations use common daily return observations.}
\label{tab:characteristics}
\footnotesize
\begin{tabular}{@{}l@{\hspace{4pt}}r@{}}
\toprule
Measure & Estimate \\
\midrule
Usable PC1 observations & 2170 \\
Active names: median [minimum, maximum] & 472 [431, 498] \\
Positive weights: median [minimum] & 100.00\% [98.80\%] \\
Explained share: median [min., max.] & 29.30\% [16.87\%, 42.88\%] \\
Top-ten weight: median & 4.06\% \\
Gross short exposure: maximum & 0.0273\% \\
Return correlation with SPY & 0.9251 \\
Return correlation with RSP & 0.9930 \\
\bottomrule
\end{tabular}
\end{table}

The leading component captures a median 29.30\% of estimated second-moment variation, ranging from 16.87\% to 42.88\%. Its return correlation is 0.9251 with SPY and 0.9930 with RSP. Together, broad participation, substantial captured variation and strong benchmark association support a market interpretation. The especially close RSP relationship motivates testing whether PC1 improves hedging beyond an equal-weighted market benchmark.

\begin{figure*}[t]
\centering
\includegraphics[width=\textwidth]{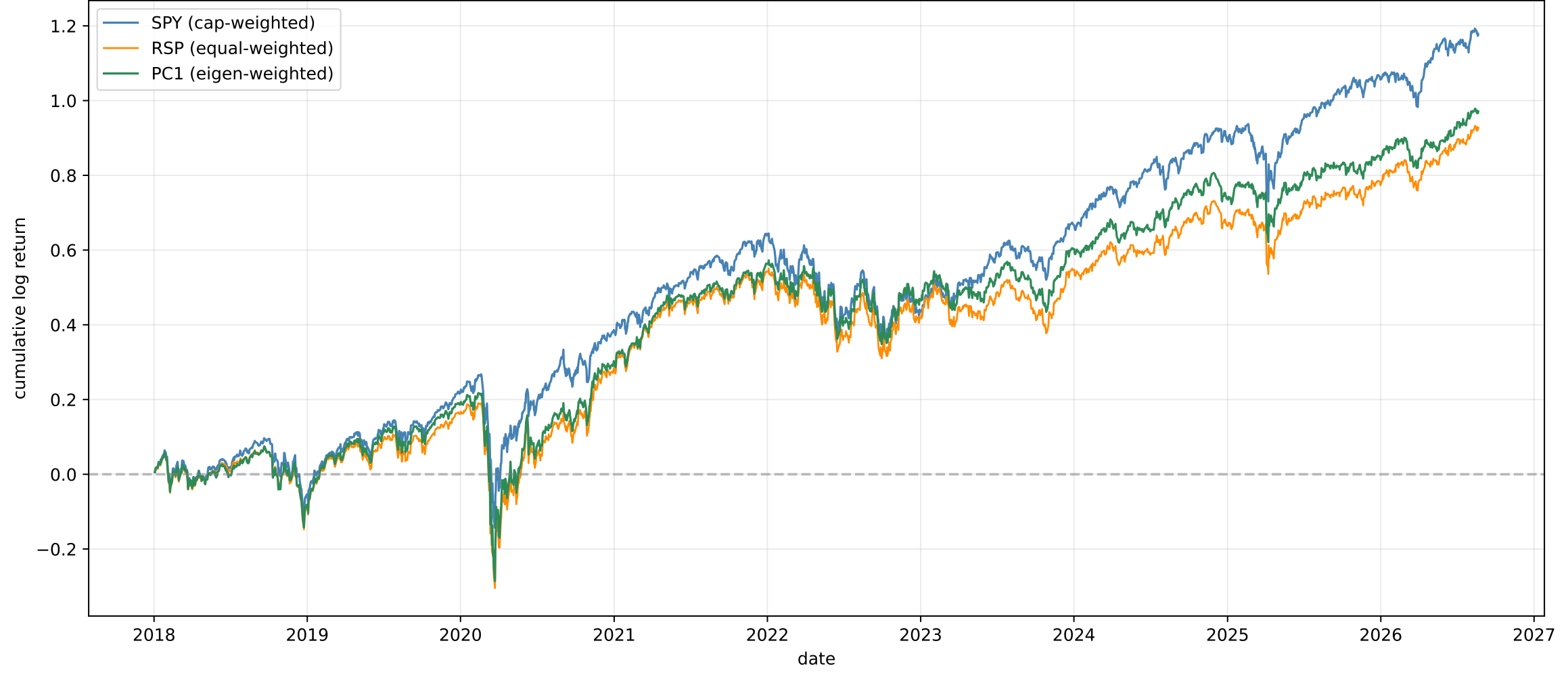}
\caption{Cumulative log returns of PC1, SPY and RSP. Daily log returns accumulated over 3 January 2018--21 August 2026. The ordinate is log compounded growth, not portfolio value in dollars. The PC1 series uses sum-to-one weights without volatility matching or added trading costs. ETF series use adjusted-price returns.}
\label{fig:cumulative}
\end{figure*}

These results indicate a broad common market direction rather than a component dominated by a small group of stocks. They do not establish recovery of a unique latent factor. Figure~\ref{fig:cumulative} shows the cumulative return histories on a common log scale. It provides a visual description of the constructed series; the hedge comparisons below provide the evidence on risk removal.

\subsection{Portfolio variance reduction}\label{sec:vr}

PC1 achieves greater mean variance reduction than SPY in every tested configuration (Table~\ref{tab:vr}). For 50-stock Dirichlet portfolios, the respective reductions are 92.17\% and 79.03\%, a difference of 13.14 percentage points. RSP achieves 91.93\%, leaving a much smaller PC1 advantage of 0.24 percentage points.

\begin{table}[t]
\centering
\caption{Mean variance reduction (\%). For each configuration, 1,000 portfolios are generated. Means use portfolios with valid variance-reduction estimates under the observation requirements in Section~\ref{sec:measures}. ``Both'' denotes the identical single-stock results under both weighting schemes; it is one shared experiment. Percentage reductions use paired portfolio/residual dates within each hedge.}
\label{tab:vr}
\footnotesize
\begin{tabular}{@{}lcccc@{}}
\toprule
Weighting & Stocks & SPY & RSP & PC1 \\
\midrule
Both & 1 & 30.80 & 35.45 & 35.72 \\
\midrule
Dirichlet & 10 & 63.52 & 73.69 & 74.03 \\
Dirichlet & 20 & 71.94 & 83.63 & 83.93 \\
Dirichlet & 50 & 79.03 & 91.93 & 92.17 \\
\midrule
Liquidity-tilted & 10 & 61.17 & 66.45 & 66.31 \\
Liquidity-tilted & 20 & 70.69 & 75.53 & 75.16 \\
Liquidity-tilted & 50 & 80.93 & 83.65 & 82.99 \\
\bottomrule
\end{tabular}
\end{table}

The comparison changes under liquidity tilting. For 50-stock portfolios, PC1 reduces variance by 82.99\%, versus 80.93\% for SPY and 83.65\% for RSP. RSP also exceeds PC1 at sizes 10 and 20 under this scheme. Thus, the advantage over SPY persists in mean variance reduction, while PC1 does not uniformly improve on RSP.

Table~\ref{tab:vrpct} reports the 10th and 90th percentiles of variance reduction across the sampled portfolios.

\begin{table}[t]
\centering
\caption{Variance reduction: 10th and 90th percentiles (\%). For each configuration, 1,000 portfolios are generated. Percentiles use portfolios with valid variance-reduction estimates under the observation requirements in Section~\ref{sec:measures}. Negative variance reductions, where present, are retained.}
\label{tab:vrpct}
\footnotesize
\begin{tabular}{@{}lcccccccc@{}}
\toprule
 & & \multicolumn{2}{c}{SPY} & \multicolumn{2}{c}{RSP} & \multicolumn{2}{c}{PC1} \\
\cmidrule(lr){3-4} \cmidrule(lr){5-6} \cmidrule(l){7-8}
Weighting & $n$ & 10th & 90th & 10th & 90th & 10th & 90th \\
\midrule
Both & 1 & 15.95 & 46.39 & 18.07 & 53.03 & 19.15 & 54.33 \\
\midrule
Dirichlet & 10 & 53.41 & 73.03 & 63.97 & 81.75 & 63.85 & 82.54 \\
Dirichlet & 20 & 64.99 & 78.94 & 78.05 & 88.25 & 77.84 & 88.76 \\
Dirichlet & 50 & 74.25 & 83.78 & 89.49 & 93.94 & 89.26 & 94.49 \\
\midrule
Liquidity-tilted & 10 & 48.70 & 72.53 & 48.17 & 79.31 & 47.13 & 79.68 \\
Liquidity-tilted & 20 & 61.38 & 79.55 & 59.74 & 85.76 & 58.13 & 86.26 \\
Liquidity-tilted & 50 & 74.12 & 87.34 & 70.28 & 91.43 & 67.52 & 91.82 \\
\bottomrule
\end{tabular}
\end{table}

Table~\ref{tab:sqres} reports the paired absolute squared-residual comparisons and their bootstrap confidence intervals.

\begin{table*}[t]
\centering
\caption{Squared-residual differences and bootstrap significance. Point estimates and 95\% block-bootstrap intervals, expressed in units of $10^{-6}$ squared daily decimal returns. Negative values favour PC1. Intervals are unadjusted; these are tests of squared residuals, not of percentage variance reduction.}
\label{tab:sqres}
\footnotesize
\begin{tabular}{@{}lcrcl@{\hspace{14pt}}rcl@{}}
\toprule
 & & \multicolumn{3}{c}{PC1 $-$ SPY} & \multicolumn{3}{c}{PC1 $-$ RSP} \\
\cmidrule(lr){3-5} \cmidrule(l){6-8}
Weighting & Stocks & Estimate & 95\% interval & Result & Estimate & 95\% interval & Result \\
\midrule
Both & 1 & $-24.007$ & [$-34.716$, $-16.012$] & PC1 lower & $+1.949$ & [$-0.160$, $+3.484$] & Not significant \\
\midrule
Dirichlet & 10 & $-23.811$ & [$-31.517$, $-17.858$] & PC1 lower & $-0.110$ & [$-1.511$, $+0.677$] & Not significant \\
Dirichlet & 20 & $-23.747$ & [$-31.498$, $-17.931$] & PC1 lower & $-0.260$ & [$-1.667$, $+0.553$] & Not significant \\
Dirichlet & 50 & $-23.618$ & [$-31.252$, $-17.947$] & PC1 lower & $-0.302$ & [$-1.686$, $+0.502$] & Not significant \\
\midrule
Liquidity-tilted & 10 & $-8.843$ & [$-13.923$, $-4.899$] & PC1 lower & $+1.579$ & [$+0.628$, $+2.390$] & Benchmark lower \\
Liquidity-tilted & 20 & $-6.674$ & [$-11.197$, $-3.249$] & PC1 lower & $+1.565$ & [$+0.567$, $+2.349$] & Benchmark lower \\
Liquidity-tilted & 50 & $-1.355$ & [$-4.091$, $+1.020$] & Not significant & $+1.699$ & [$+0.772$, $+2.518$] & Benchmark lower \\
\bottomrule
\end{tabular}
\end{table*}

\paragraph{Statistical significance of squared-residual differences.} The squared-residual comparisons favour PC1 over SPY significantly at every size under Dirichlet weighting and at sizes 1, 10 and 20 under liquidity tilting. At size 50 under liquidity tilting, the interval includes zero. No PC1--RSP squared-residual difference is significant under Dirichlet weighting; RSP has significantly smaller squared residuals for all multi-stock liquidity-tilted portfolios. The slight PC1 advantage in mean percentage variance reduction at $n = 1$ coexists with a positive, statistically inconclusive squared-residual difference against RSP: the two statistics use different scaling and aggregation.

The PC1--RSP ranking varies with portfolio weighting. Liquidity tilting changes both portfolio exposures and concentration, and the experiment does not isolate which of these differences explains RSP's advantage.

\subsection{Residual cross-sectional correlation}\label{sec:corr}

All three hedges substantially reduce average portfolio co-movement (Table~\ref{tab:corr}). For 50-stock Dirichlet portfolios, mean correlation falls from 0.9335 unhedged to 0.6840 under SPY, 0.1593 under RSP and 0.1244 under PC1. PC1 has the lowest descriptive mean correlation across the Dirichlet configurations. Under liquidity tilting, it is close to RSP at size 10 but leaves higher mean correlation at sizes 20 and 50.

\begin{table}[t]
\centering
\caption{Mean portfolio correlation before and after hedging. Descriptive panels use 250 portfolios and instrument-specific filtering. Retained date counts are 2,042 for single-stock portfolios and 2,044 for the larger portfolios. Matching counts alone do not establish identical selected observations.}
\label{tab:corr}
\footnotesize
\begin{tabular}{@{}lccccc@{}}
\toprule
Weighting & $n$ & Unhedged & SPY & RSP & PC1 \\
\midrule
Both & 1 & 0.3487 & 0.0839 & 0.0133 & 0.0046 \\
\midrule
Dirichlet & 10 & 0.7443 & 0.3164 & 0.0397 & 0.0204 \\
Dirichlet & 20 & 0.8456 & 0.4599 & 0.0709 & 0.0496 \\
Dirichlet & 50 & 0.9335 & 0.6840 & 0.1593 & 0.1244 \\
\midrule
Liquidity-tilted & 10 & 0.6723 & 0.1757 & 0.0295 & 0.0293 \\
Liquidity-tilted & 20 & 0.7663 & 0.2172 & 0.0685 & 0.0767 \\
Liquidity-tilted & 50 & 0.8546 & 0.2788 & 0.1572 & 0.1777 \\
\bottomrule
\end{tabular}
\end{table}

Table~\ref{tab:corrdiff} makes the correlation comparison on common portfolios and dates and applies the same sign convention as Table~\ref{tab:sqres}. Its entries are differences in average correlation coefficients, not percentage changes in correlation.

\begin{table*}[t]
\centering
\caption{Residual-correlation differences and bootstrap significance. Point differences and 95\% block-bootstrap intervals use jointly aligned portfolios and dates. Negative values favour PC1. Each comparison uses 250 portfolios: 2,042 retained dates for $n = 1$ and 2,044 otherwise. Intervals are unadjusted for multiple comparisons.}
\label{tab:corrdiff}
\footnotesize
\begin{tabular}{@{}lcrcl@{\hspace{14pt}}rcl@{}}
\toprule
 & & \multicolumn{3}{c}{PC1 $-$ SPY} & \multicolumn{3}{c}{PC1 $-$ RSP} \\
\cmidrule(lr){3-5} \cmidrule(l){6-8}
Weighting & Stocks & Estimate & 95\% interval & Result & Estimate & 95\% interval & Result \\
\midrule
Both & 1 & $-0.07933$ & [$-0.08886$, $-0.06926$] & PC1 lower & $-0.00875$ & [$-0.01177$, $-0.00608$] & PC1 lower \\
\midrule
Dirichlet & 10 & $-0.29593$ & [$-0.32302$, $-0.26740$] & PC1 lower & $-0.01932$ & [$-0.03406$, $-0.00716$] & PC1 lower \\
Dirichlet & 20 & $-0.41032$ & [$-0.44015$, $-0.37830$] & PC1 lower & $-0.02131$ & [$-0.04937$, $+0.00295$] & Not significant \\
Dirichlet & 50 & $-0.55962$ & [$-0.58490$, $-0.53100$] & PC1 lower & $-0.03491$ & [$-0.09152$, $+0.01750$] & Not significant \\
\midrule
Liquidity-tilted & 10 & $-0.14635$ & [$-0.16361$, $-0.12791$] & PC1 lower & $-0.00017$ & [$-0.00815$, $+0.00712$] & Not significant \\
Liquidity-tilted & 20 & $-0.14048$ & [$-0.16279$, $-0.11487$] & PC1 lower & $+0.00827$ & [$-0.00751$, $+0.02224$] & Not significant \\
Liquidity-tilted & 50 & $-0.10107$ & [$-0.12829$, $-0.06798$] & PC1 lower & $+0.02054$ & [$-0.00343$, $+0.04252$] & Not significant \\
\bottomrule
\end{tabular}
\end{table*}

\paragraph{Statistical significance of residual-correlation differences.} On jointly aligned panels, PC1 produces significantly lower mean residual correlation than SPY in every tested configuration (Table~\ref{tab:corrdiff}). For 50-stock liquidity-tilted portfolios, the difference is $-0.10107$, with a 95\% interval of $[-0.12829, -0.06798]$, even though the squared-residual comparison for that configuration is inconclusive. The two outcomes therefore provide complementary information about individual portfolio fluctuations and residual co-movement.

Against RSP, significant reductions in mean residual correlation occur for single-stock portfolios and 10-stock Dirichlet portfolios. The intervals include zero for the remaining configurations. In particular, the positive PC1--RSP point differences for 20- and 50-stock liquidity-tilted portfolios do not establish a statistically significant increase in mean residual correlation.

Across the two outcomes, the evidence is more consistent against SPY than against RSP. PC1's broad positive participation and close return association with RSP accompany substantial improvements over SPY, but its comparison with RSP depends on portfolio weighting and the performance measure. This pattern is consistent with a role for weighting, while the experiment does not isolate weighting as the cause of the differences.

%% ======================================================================
\section{Discussion and conclusion}\label{sec:discussion}

This study examined whether an EWMA-based principal eigenportfolio improves beta hedging relative to SPY and RSP. The component exhibits broad market participation and close association with RSP. PC1 achieves greater mean variance reduction and lower average residual correlation than SPY across the tested configurations. Its comparison with RSP is smaller and mixed: PC1 slightly improves mean variance reduction for Dirichlet portfolios, whereas RSP performs better for multi-stock liquidity-tilted portfolios. Statistical comparisons support the SPY-relative advantage more consistently than an advantage over RSP.

The central implication is that extracting a convincing market component and constructing a better hedge are distinct achievements. Broad participation, captured variation and benchmark association support PC1's interpretation as a market component; they do not establish an incremental hedging benefit. RSP changes how the SPY comparison should be understood. For 50-stock Dirichlet portfolios, PC1's mean variance reduction exceeds SPY's by 13.14 percentage points, but exceeds RSP's by only 0.24 percentage points, with an inconclusive squared-residual comparison against RSP. Against SPY alone, the improvement might be attributed to statistical extraction. The RSP comparison shows that an existing equal-weight instrument already delivers much of that benefit. Benchmark selection therefore affects the substantive conclusion about what PCA adds, rather than merely providing another performance comparison.

The different variance-reduction rankings across portfolio designs suggest that hedge quality depends on the relationship between hedge and target. A component that represents broad market movement well need not capture every portfolio's common exposure equally well. This shifts attention from identifying a universally best market proxy to identifying where a particular proxy adds hedge value. Weighting and exposure differences are plausible contributors to the observed pattern, but the experiment does not isolate them: PC1 and RSP differ in allocations and updating, while liquidity tilting changes target exposures and concentration. The two performance measures also matter. PC1 significantly lowers residual correlation relative to SPY in every configuration, even though its squared-residual advantage is inconclusive for 50-stock liquidity-tilted portfolios. Reducing shared residual movement and reducing the magnitude of residual risk are related, but empirically distinct, benefits.

These conclusions are conditional on the studied universe, period, portfolio designs and estimation settings. Historical membership reconstruction improves universe coverage but does not eliminate selection arising from unavailable stock histories. The fixed parameters do not establish optimal horizons or universal hedge rankings. Bootstrap intervals describe temporal uncertainty conditional on the realised portfolio and hedge histories and are unadjusted for multiple comparisons. Additional trading, financing and borrowing costs are excluded, so the results establish gross statistical hedge effectiveness rather than a net implementation benefit.

Further work could clarify the mechanism through benchmark comparisons that more closely control weighting, concentration and other exposure differences. Evaluating turnover and realistic implementation costs would then help determine whether the incremental risk reduction justifies maintaining the constructed portfolio.

EWMA-PC1 thus provides a useful market representation and an effective alternative to SPY in the tested setting. Its incremental value depends on the benchmark and target portfolio: the evidence supports evaluating data-derived hedges against simple alternatives, rather than inferring superiority from market resemblance alone.

%% ======================================================================
\section*{Data availability}

Daily adjusted prices were obtained from Yahoo Finance, and historical S\&P 500 membership was reconstructed from Wikipedia constituent records, as described in Section~\ref{sec:data}. The cached snapshots used in the reported run are not redistributed with this paper.

\section*{Disclosure of AI use}

Generative AI tools were used to assist with editing the manuscript text and with checking references against their published records. The authors reviewed all text, code and results and take full responsibility for the content of this paper.

%% ======================================================================

\end{document}